\documentclass[10pt,a4paper,notitlepage]{revtex4-1}
\usepackage{graphicx}
\usepackage{amsmath}
\usepackage{subcaption}
\usepackage{color}
\usepackage{fullpage}

\begin{document}
\title{Combined complex Doppler and Cherenkov effect in left-handed metamaterials}
\author{D. Ziemkiewicz}
\email{david.ziemkiewicz@utp.edu.pl}
\author{S. Zieli\'nska-Raczy\'nska}
\affiliation{Institute of Mathematics and Physics, UTP University of Science and Technology, Al. Kaliskiego 7, 85-798 Bydgoszcz, Poland.}

\begin{abstract}
We derive the formula of the complex Doppler shift in a two-dimensional, dispersive metamaterial and we show that a moving, monochromatic radiation source generates multiple frequency modes. The role of the group velocity is stressed and the Doppler shifted radiation field exhibits features of the Cherenkov effect. The presented theory is also applicable to the case of a moving, non-oscillating charge and explains many peculiar characteristics of the Cherenkov radiation in left-handed metamaterials such as the backward direction of power emission, the constant radiation angle and the lack of velocity threshold.
\end{abstract}
\maketitle 
\section{Introduction}
The negative refraction index media, first considered by Veselago \cite{Veselago}, and recently realized in the form of metamaterials~\cite{Smith,Shelby} exhibit many fascinating properties \cite{Veselago2,Pendry1} and potential applications such as imaging beyond the diffraction limit \cite{Pendry2}. In such materials, commonly called left-handed media (LHM) due to left-handed triada of vectors $\vec E$, $\vec H$ and $\vec k$  of propagating waves, many usual optical phenomena become reversed, however some of them are unreversed, just to mention the so-called rotational Doppler effect \cite{Luo} as well as the spin Hall effect in  LHM \cite{Luo2}. The class of reversed phenomena includes
refraction \cite{Pendry1,Shelby}, Goos - H\"{a}nchen shift~\cite{GHshift} and the Doppler effect \cite{Veselago}. The reversed Doppler shift has been observed in a variety of systems such as magnetic thin film~\cite{Spinwave}, photonic crystals~\cite{Dop_Opt}, transmission lines~\cite{Seddon} and acoustical metamaterials~\cite{Sound_Lee}. In dispersive media, the so-called complex Doppler effect occurs, when a monochromatic source generates wave modes of multiple frequencies \cite{Ziemkiewicz}.

Another example of reversed phenomena in LHM is the Cherenkov effect \cite{Veselago}. This radiation, which was discovered  experimentally in 1934 by Cherenkov \cite{Cherenkov} and later described theoretically  by Frank and Tamm \cite{Tamm} occurs when a charged particle moves at a speed exceeding the phase velocity of the waves in a medium  and has been extensively applied for particle detection. In usual, right-handed materials, the angle of radiation is smaller than $\pi/2$, so that the energy is emitted in the forward direction.
One of the important feature of negative index materials is a negative value of the phase velocity so the Cherenkov effect is reversed and the waves are emitted in backward direction \cite{Veselago}. It has been observed experimentally by Xi \emph{et al.}\cite{Xi}. The Cherenkov  radiation in LHM has been studied by Lu \emph{et al.} \cite{Lu} for both lossless and lossy cases and they found that maintaining a forward $k$ vector of radiation in negative index media exhibits a backward emission. An elegant review of Cherenkov radiation in photonic crystals describing a backward-pointing radiation cone and backward direction of emission was presented by Luo \emph{et al.} \cite{Luo_Phot}. More recently in \cite{Chen} it was pointed out that the reversed Cherenkov radiation  has a distinct advantage which allows the photons and charged particle to naturally separate in opposite directions, minimizing their interference. Nowadays there is a considerable interest in developing a new class of velocity-sensitive particle detectors based on left-handed materials \cite{Lu} which can be used to enhance the radiation \cite{Duan} or control its emission angle \cite{Ginis}.

Here we present a simple, unified analytical description of the Doppler and Cherenkov phenomena in an idealized, dispersive medium. The description of the complex Doppler effect presented in our previous paper \cite{Ziemkiewicz} is extended to the two-dimensional case. It is shown that a moving, monochromatic source generates two distinct frequency modes and the generated field exhibits features of both the Doppler and the Cherenkov effects.
Starting from the first principles, we are able to explain the angular distribution of the energy for the fast and slow frequency modes of the two-dimensional, complex Doppler effect in LHM. Moreover, the reversed Cherenkov radiation emerges as a particular case of the presented theory as the source with the frequency $\omega_0=0$ is shown to be formally equivalent to a moving charge. The intrinsic feature of the reversed Cherenkov effect in the considered LHM is the lack of a source velocity threshold above which this phenomenon occurs. This fact is explained in a general way and is shown to have a different origin than in the case of periodic media such as nanowire structures \cite{Fernandes}.

The theoretical findings are confirmed by numerical simulations based on the Finite Difference Time Domain (FDTD) method which is a common tool in the study of electromagnetic wave propagation in various media \cite{Veselago2,KS_Taflove}. The performed numerical simulations are in agreement with our theoretical predictions based on the first principle approach.

This paper is organized as follows. In section 2 we present the theory of the complex Doppler effect in two-dimensional LHM, section 3 is devoted to the description of the simulation setup. A spatial structure of fields emitted by a moving source characterized by nonzero and zero frequency, describing the complex Doppler and the reversed Cherenkov effect respectively, is discussed in section 4. Conclusions are drawn in section 5. Finally, in Appendix A we prove that the Cherenkov cone angle varies only within small interval and is almost insensible to the source velocity.

\section{Theory}
\begin{figure}[ht!]
\begin{center}
    \includegraphics[scale=0.6]{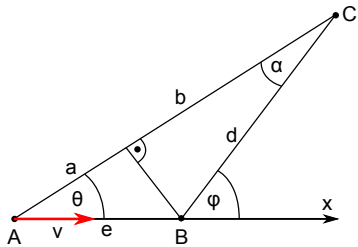}
  \caption{Source of radiation moving with velocity $v$ emits two subsequent wave fronts at points A and B, towards observer C.}\label{Doppler_rys}
\end{center}
\end{figure}

Consider an observer C receiving wavefronts emitted by a point source moving at a velocity $\vec v$, where the initial angle between source-observer line and $\vec v$ is $\theta$, as shown on the Fig. \ref{Doppler_rys}. The radiation source has a nominal frequency $\omega_0 = \frac{2\pi}{T_0}$ in its own reference frame.
    Assuming that the first wavefront was emitted at $t=0$, the time instant in which  it will be registered is
    \begin{equation}
    t_1 = \frac{a+b}{v_p(\omega)} = \frac{n(\omega)(a+b)}{c},
    \end{equation}
where $v_p(\omega)$ is the phase velocity in the medium, and $n(\omega)$ is the refraction index. These quantities are observed in the reference frame of the detector and are functions of the detected frequency $\omega$. The second wavefront will arrive at the receiver at the time instant
    \begin{equation}
    t_2 = \frac{T_0}{\sqrt{1-\beta^2}} + \frac{n(\omega)d}{c},
    \end{equation}
where $\beta = v/c$. The period measured by the observer will be
\begin{equation}
T = \left|\frac{T_0}{\sqrt{1-\beta^2}} - \frac{n(\omega)}{c}(a+b-d)\right|,
\end{equation}
where the absolute value was used to ensure that the period is a positive quantity. In a typical case, the distance $e$ traveled during a single period is negligible. Therefore, the angle $\alpha \approx 0$ and $b \approx d$. By using the relation
\begin{equation}
a = e \cos \theta = v\frac{T_0}{\sqrt{1-\beta^2}}\cos \theta,
\end{equation}
one obtains
\begin{equation}\label{DOPP1}
\omega = \omega_0 \left|\frac{\sqrt{1-\beta^2}}{1 - n(\omega)\beta \cos \theta}\right|,
\end{equation}
similar to the result obtained in \cite{Bazhanova},
where in our case the absolute value is necessary to ensure that the frequency is a positive quantity. The above relation is implicit and has multiple solutions in a dispersive medium, leading to the so-called the complex Doppler effect \cite{Frank}, where a monochromatic source generates multiple detected frequencies.

To solve the Eq. \ref{DOPP1}, one has to specify the dispersion relation of the medium. A remarkably simple description discussed by Veselago \cite{Veselago} in the context of the left-handed media is given by the Drude model, with the permittivity and permeability given by $\epsilon(\omega)= 1 - \frac{\omega_{pe}^2}{\omega^2}$ and $\mu(\omega)= 1 - \frac{\omega_{pm}^2}{\omega^2}$ respectively, where $\omega_{pe}$ and $\omega_{pm}$ depend on the metamaterial structure \cite{Pendry_6}. The proposed relations are applicable to a wide range of metamaterials \cite{Veselago2,Pendry_6}. For further simplification, one can assume $\omega_{pe} = \omega_{pm} = \omega_p$, so that the dispersion relation for the refraction index $n$ is
\begin{equation} \label{modDRU}
n(\omega) = 1 - \frac{\omega_p^2}{\omega^2}.
\end{equation}
With the above dispersion model, from Eq. (\ref{DOPP1}) one obtains two explicit solutions of the following form
\begin{eqnarray}\label{DOPP2}
\omega_1(\beta,\theta) = \frac{\omega_0\sqrt{1-\beta^2} + \sqrt{(1-\beta^2)\omega_0^2 - 4x(1-x)\omega_p^2}}{2(1-x)},\nonumber\\
\omega_2(\beta,\theta) = \frac{\omega_0\sqrt{1-\beta^2} - \sqrt{(1-\beta^2)\omega_0^2 - 4x(1-x)\omega_p^2}}{2(1-x)},
\end{eqnarray}
where
\begin{equation}
x = \beta\cos\theta.
\end{equation}
The important property of the Eqs. (\ref{DOPP2}) is the range of parameters where the frequencies $\omega_{i}$ are real, indicating non-evanescent wave modes detectable in the far field of the source. By taking, for simplicity, $\omega_0^2 = \frac{1}{2}\omega_p^2$ so that $n(\omega_0)=-1$, one obtains
\begin{equation}
\cos \theta \leq \frac{2 - \sqrt{2 + 2\beta^2}}{4 \beta}
\end{equation}
so that for sufficient source velocity, no propagating modes will be observed for the angle $\theta$ smaller than some limit value $0 \geq \theta_0 \geq \pi/2$. By inspecting the implicit Eq. (\ref{DOPP1}), one can see that for the negative value of the refraction index, the Doppler shift is reversed, so that the frequency of waves emitted in the forward direction ($\theta < \pi/2$) will be downshifted. In the considered model, the frequency of the upshifted modes at $\theta > \pi/2$ is inherently limited; as $\omega \rightarrow \omega_p$, the refraction index $n \rightarrow 0$ and the denominator of Eq. (\ref{DOPP1}) is equal to $1$. Therefore, the frequency $\omega_p$ cannot be exceeded when $\omega_0<\omega_p$. This means that a source having frequency $\omega_0$ such that $n(\omega_0)<0$ will generate only negative phase velocity modes with $n(\omega) < 0$. 
\section{Simulation setup}
The performed simulations were based on a standard FDTD algorithm \cite{Yee} where the Maxwell's equations are solved in the time domain. A two-dimensional system with transverse magnetic field is chosen to facilitate the simulation of Cherenkov radiation \cite{Chen}; the electric field has two components in the plane of propagation $\vec E = [E_x,E_y,0]$, and the magnetic field has a single component $\vec H = [0,0,H_z]$. To model the dispersive medium, the Auxiliary Differential Equations (ADE) method is used \cite{KS_Taflove}. The implementation is a straightforward extension of the method used in our preceding work \cite{Ziemkiewicz} and is based on the time domain calculation of the polarization and magnetization in the medium. In the considered Drude model, these quantities are given by equations of motion $\ddot{P} = \omega_p^2 E$ and $\ddot{M} = \omega_p^2 H$.

The source with frequency $\omega_0$ moving at a velocity $V$ along $\hat x$ axis is modeled as a time varying current density in the form
\begin{equation}
J = J_0 \exp\left[\frac{-(x-Vt)^2}{2\sigma^2}\right]\exp\left(\frac{-y^2}{2\sigma^2}\right)\exp(-i\omega_0 t).
\end{equation}
The above representation describes an oscillating dipole oriented along $x$ axis \cite{Luo_Phot}, where the usual Dirac delta function is replaced with sufficiently narrow Gaussian width, with the width $\sigma$ significantly smaller that the source wavelength $\lambda(\omega_0)$. The only relativistic effect taken into account is the time dilation affecting the frequency $\omega_0$, which is needed to keep the consistency with the relativistic description of the Doppler shift presented above. In the performed simulations, the Lorentz contraction and other factors affecting the field amplitude can be disregarded, as the following discussion of the radiation intensity is qualitative in nature and the relativistic effects at the considered speeds are not significant.

The simulation space is divided into 300x300 grid and is surrounded by absorbing boundaries to reduce reflections. The frequency $\omega_0$ was chosen to minimize the effect of the numerical dispersion and anisotropy \cite{KS_Taflove}. This is especially important in the case of a dispersive medium described above, where the high frequency modes are limited by the finite time step, and the low frequency, short wavelength modes are constrained by the spatial step.

\section{Spatial structure of the field}
\begin{figure}[ht!]
\begin{center}
    \includegraphics[scale=0.4]{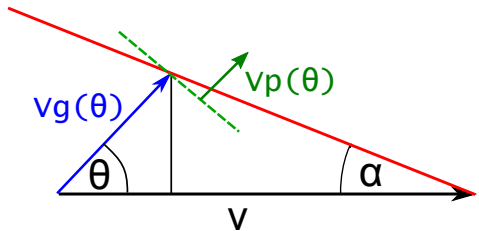}
  \caption{Radiation cone formed by a single mode emitted at a given angle $\theta$. Dashed line indicates the wavefront.}\label{fig:2}
\end{center}
\end{figure}

\begin{figure}
\begin{minipage}[b]{.5\linewidth}
\centering
\includegraphics[scale=0.37]{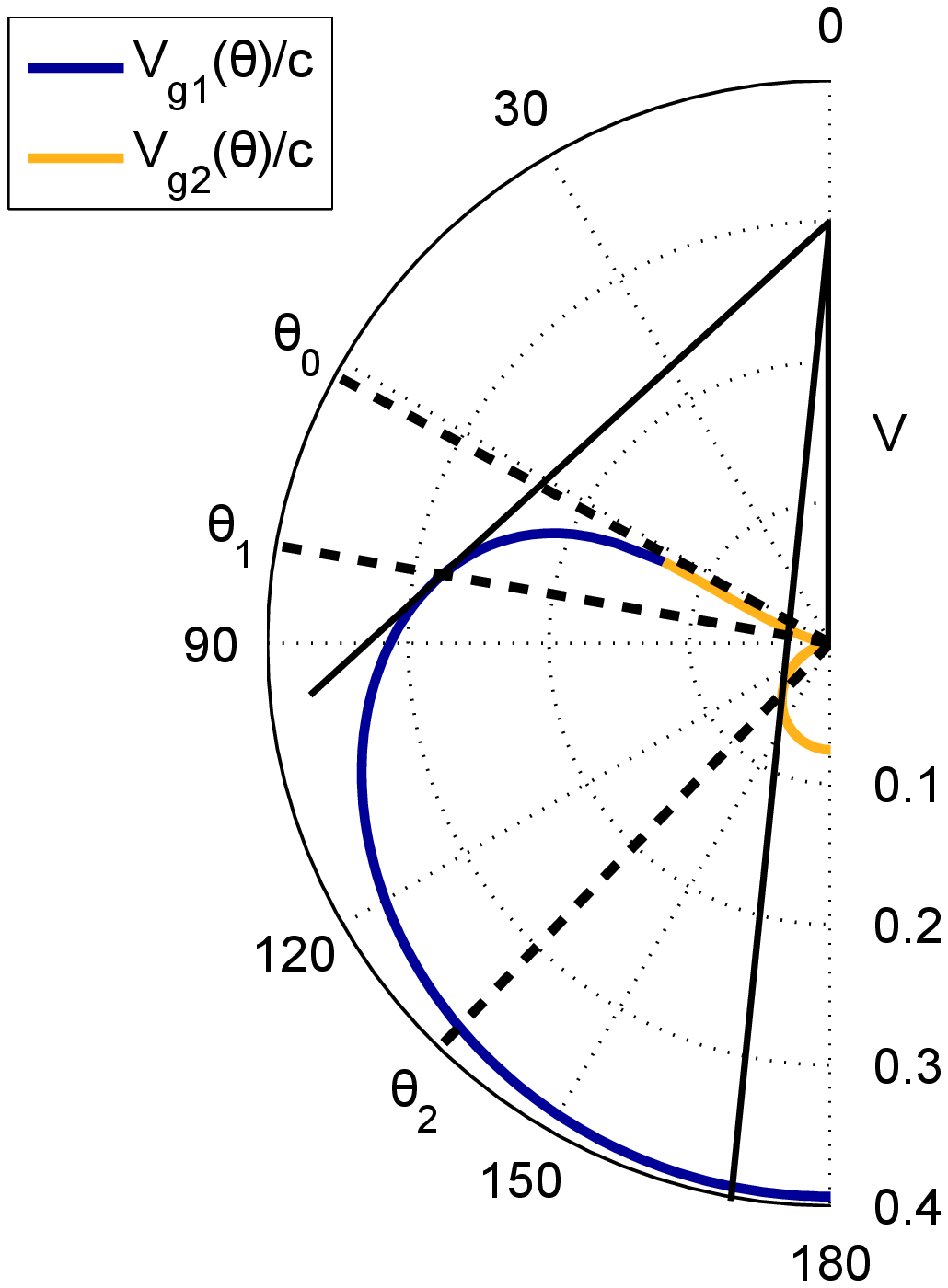}
\subcaption{}\label{fig:3a}
\end{minipage}%
\begin{minipage}[b]{.5\linewidth}
\centering
\includegraphics[scale=0.27]{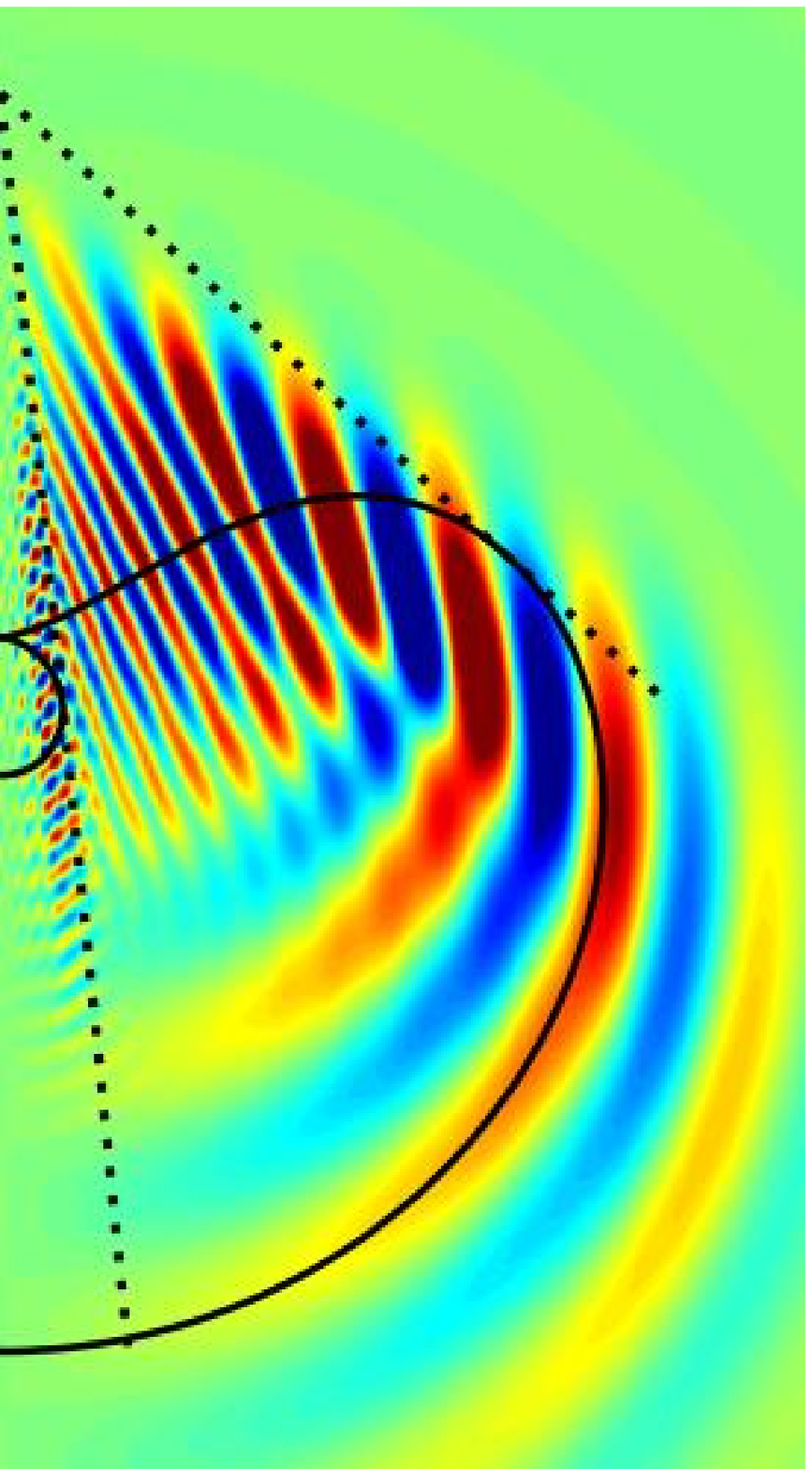}
\subcaption{}\label{fig:3b}
\end{minipage}

\caption{(a) Radiation cones of the two Doppler modes at $\beta=0.3c$ and $n(\omega_0)=-1$. The angles $\theta_1$ and $\theta_2$ mark the direction of radiation forming the cone. (b) The group velocity contour imposed on the field snapshot calculated with FDTD method.}\label{fig:3}
\end{figure}

Let's assume that the moving source emits a quasi-monochromatic wave packet with central frequency $\omega_0$. According to Eq. (\ref{DOPP2}), for any emission angle $\theta$, there are two wave modes $\omega_i(\theta)$, $i=1,2$. Therefore, the wavefronts will propagate at an angle-dependent phase velocity $V_{pi}(\theta)=V_{pi}[\omega(\theta)]$. However, their envelopes will move at a group velocity $V_{gi}(\theta) = c/\left(n(\omega) + \omega\frac{\partial n}{\partial \omega}\right)$ which is assumed to be positive. When $V_g \cos \theta < V$, the emitted field will be detectable only behind the source, forming the so-called group cone \cite{Carusotto}. In dispersive media, its apex angle may differ significantly from the usually considered wave cone formed by the wavefronts. The situation is shown on the Fig. \ref{fig:2}. Assuming that the medium is isotropic and lossless, the phase velocity will be parallel to the group velocity.
As it follows from the geometry of the~Fig.~\ref{fig:2}, the cone half apex angle $\alpha$ can be expressed as
\begin{equation}\label{rw0}
\tan \alpha = \frac{V_g(\theta) \sin \theta}{V - V_g(\theta) \cos \theta}.
\end{equation}
To deduce the whole radiation pattern of the moving source, one has to consider the wave modes for all angles $\theta$. An elegant way of presentation was pointed out in \cite{Luo_Phot} and it consists of plotting the group velocity $V_g(\theta)$ as a contour and comparing it to the source velocity. By doing this, one can visualize the distance traveled by the source and the wave envelope at some given time, recreating the spatial structure of the radiated field. An example of such construction for $\beta=0.3c$ and $n(\omega_0)=-1$ is shown on the Fig. \ref{fig:3a} and the corresponding field snapshot is presented on the Fig. \ref{fig:3b}. The solutions of Eq. (\ref{DOPP2}) define two contours $V_{gi}(\theta)$. The high frequency modes $\omega_1(\theta)$ are comparable to the nominal frequency $\omega_0$ and are characterized by relatively high group velocity $V_{g1}(\theta)$. This solution can be associated with the usual Doppler effect; the backwards moving waves ($\theta > \pi/2$) are upshifted, with their group velocity exceeding the value of $V_g(\omega_0) = c/3$. Also, the transverse Doppler shift for the waves emitted in a direction perpendicular to the source motion is caused only by the relativistic effects, so in the low velocity regime $\omega_1(\pi/2)\approx\omega_0$. For the forward moving waves, the frequency and the group velocity drops quickly as $\theta \rightarrow \theta_0$, possibly becoming much smaller than $V$ and leading to the Cherenkov-like radiation pattern. The second wave mode $\omega_2$ is characterized by small group velocity. A characteristic point of this solution is $\theta=\pi/2$, where $\omega_2 = 0$. As the angle approaches this value, the frequency becomes arbitrarily small. In a realistic scenario, the amount of the downshift is limited to the range where the dispersion model in Eq. (\ref{modDRU}) is applicable. Moreover, the low frequency modes are more affected by the absorption \cite{Ziemkiewicz}, making their detection difficult.

\begin{figure}
\begin{minipage}[b]{.5\linewidth}
\centering
\includegraphics[scale=0.21]{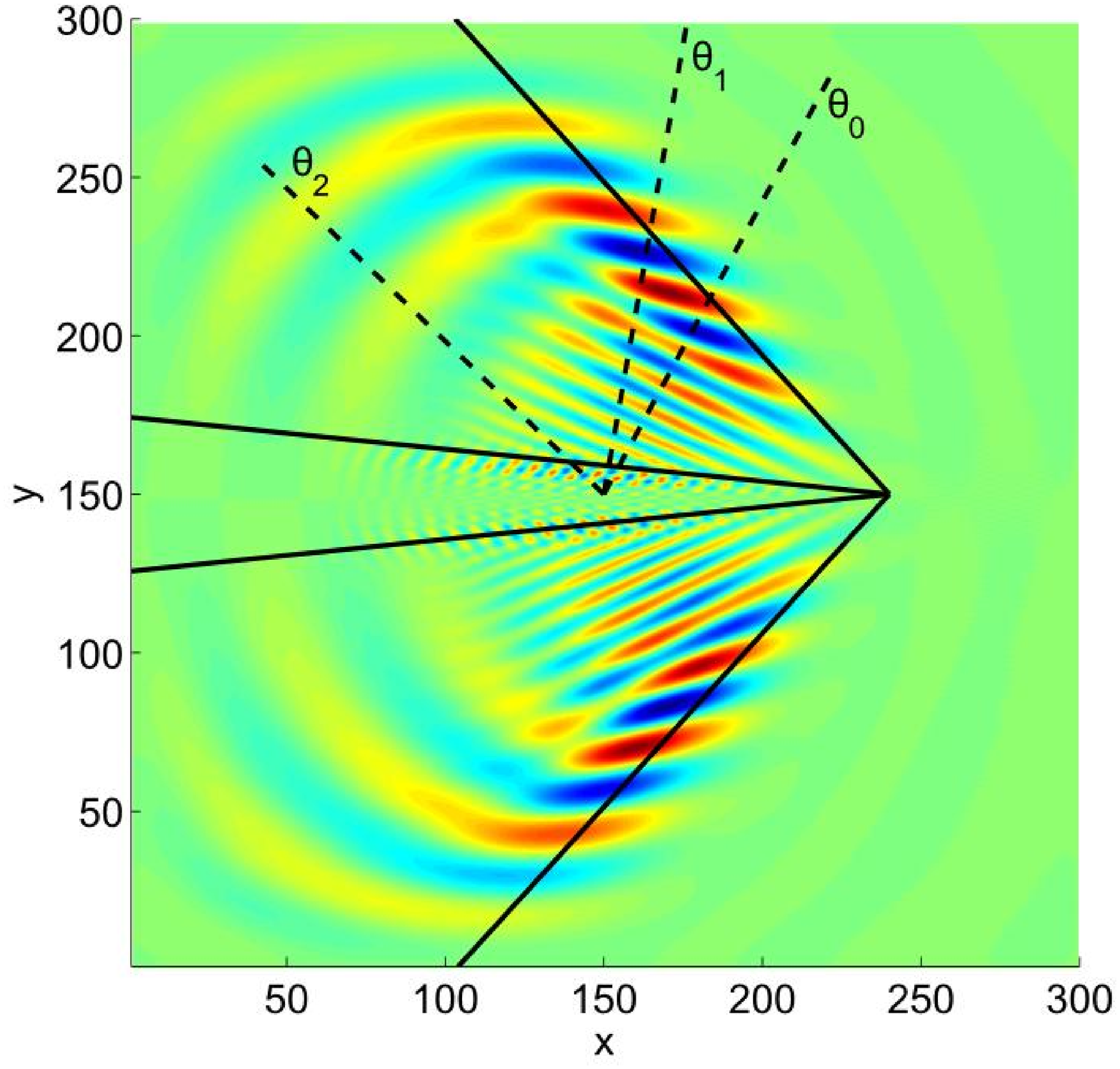}
\subcaption{}\label{fig:4a}
\end{minipage}%
\begin{minipage}[b]{.5\linewidth}
\centering
\includegraphics[scale=0.33]{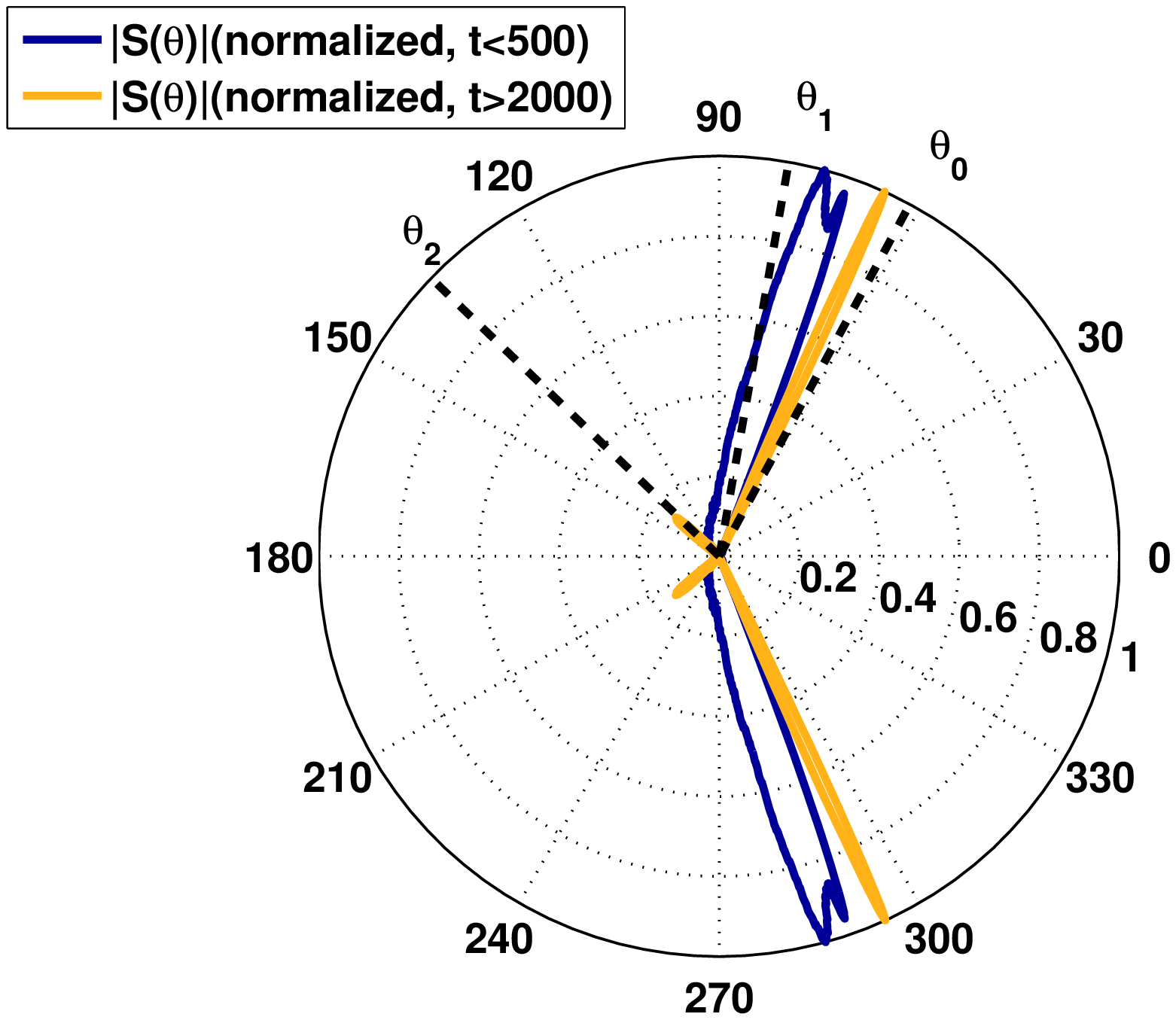}
\subcaption{}\label{fig:4b}
\end{minipage}

\begin{minipage}[b]{.5\linewidth}
\centering
\includegraphics[scale=0.33]{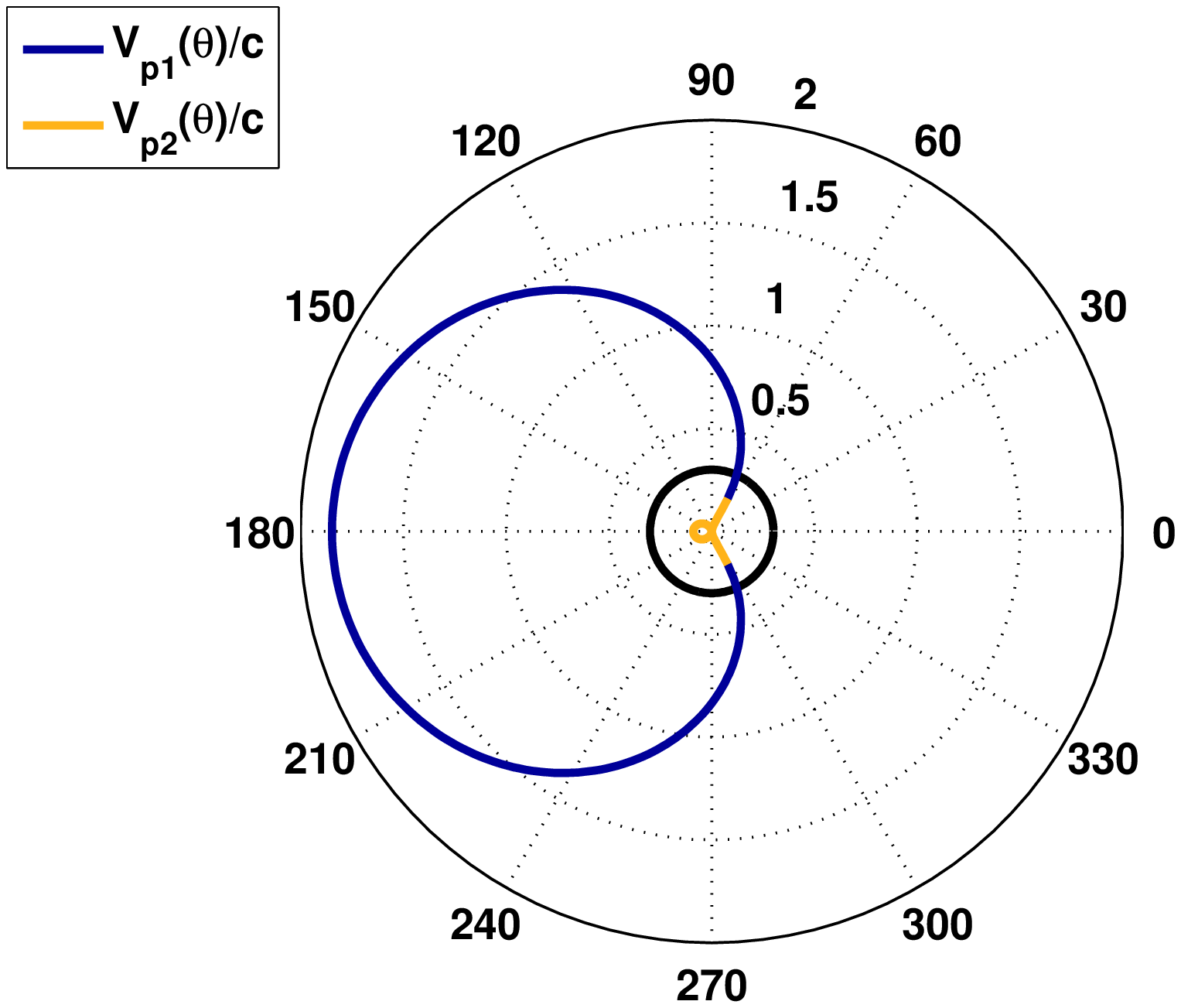}
\subcaption{}\label{fig:4c}
\end{minipage}%
\begin{minipage}[b]{.5\linewidth}
\centering
\includegraphics[scale=0.33]{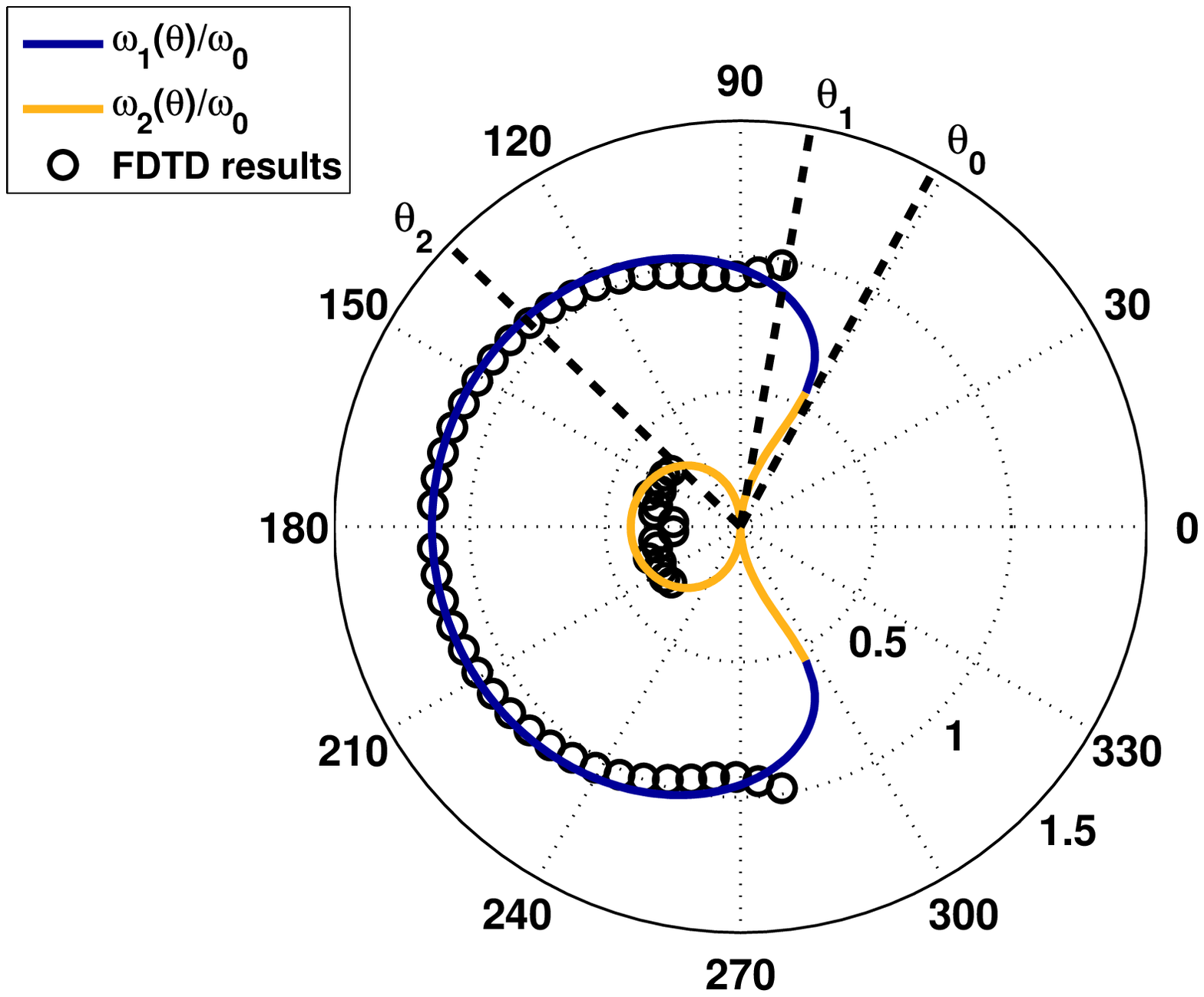}
\subcaption{}\label{fig:4d}
\end{minipage}

\begin{minipage}[b]{.5\linewidth}
\centering
\includegraphics[scale=0.21]{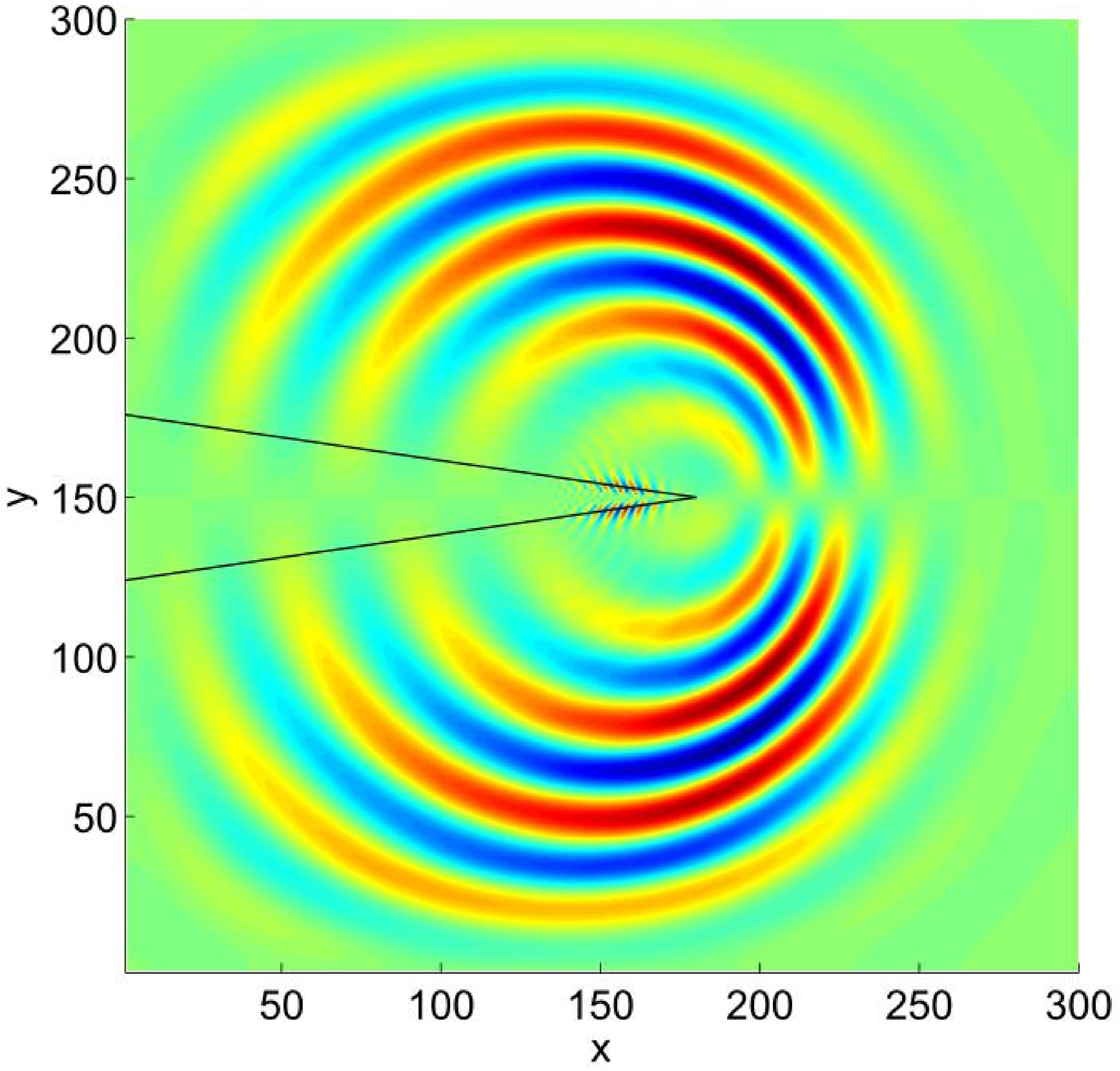}
\subcaption{}\label{fig:4e}
\end{minipage}%
\begin{minipage}[b]{.5\linewidth}
\centering
\includegraphics[scale=0.33]{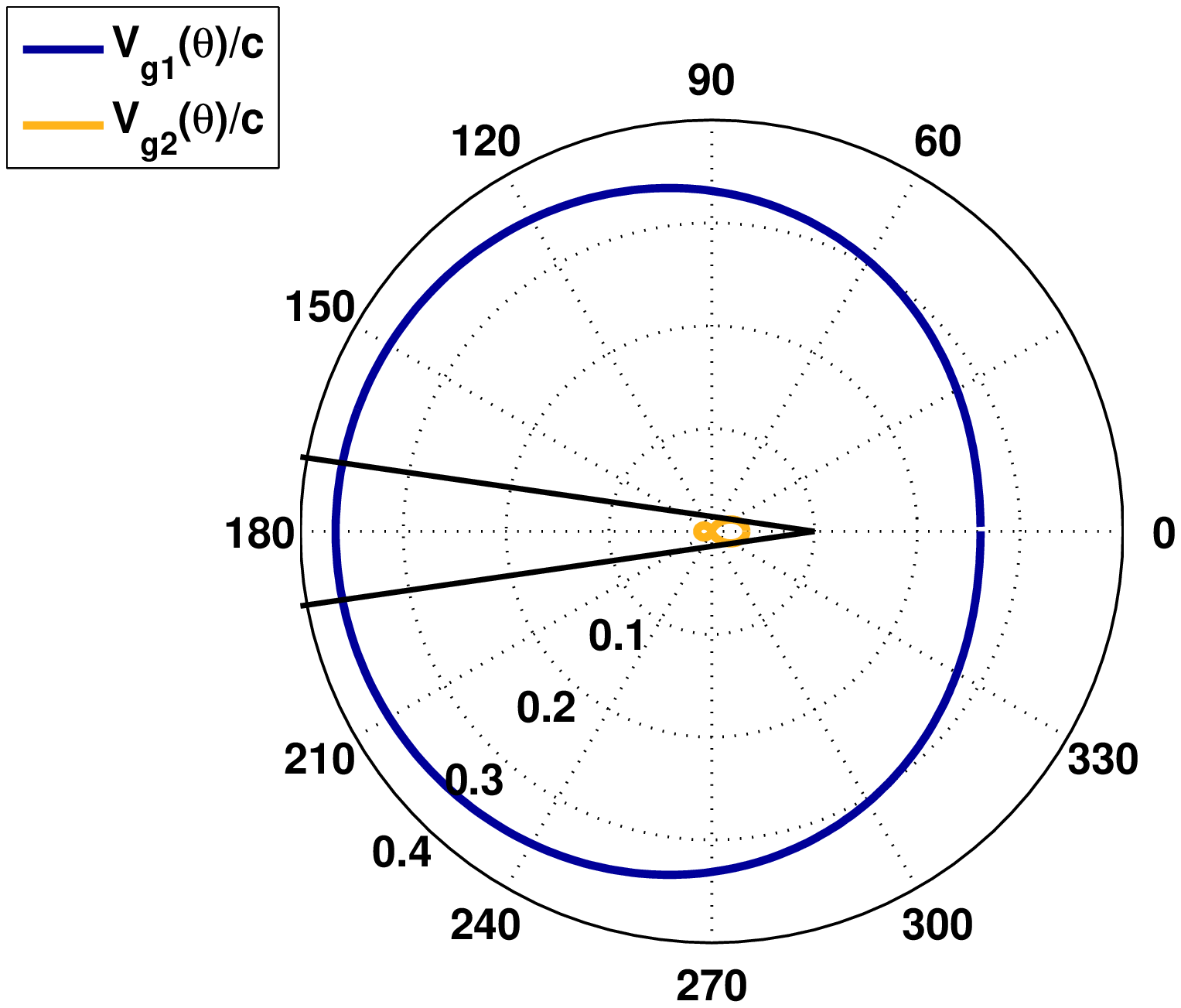}
\subcaption{}\label{fig:4f}
\end{minipage}

\caption{Simulation results for the complex Doppler effect at $\beta = 0.3$ and $n(\omega_0)=-1$. (a) Field snapshot at t=500. The cones clearly divide the space into distinct areas. (b) Angular distribution of energy for the fast modes centered around $\theta_1$ and the low frequency modes at $\theta_0$ and $\theta_2$. (c) The phase velocity contour. The radius of the black circle is equal to the source speed. (d) The frequency as a function of angle measured in the area where the field is monochromatic. (e) Field snapshot for $\beta=0.1$. (f) The group velocity contour for $\beta=0.1$.}\label{fig:4}
\end{figure}

The space taken by the radiated field is limited by the maximum value of cone angle $\alpha$ measured between the source velocity and a line tangent to the group velocity contour (Fig. \ref{fig:3a}). Its value can be determined by solving $\partial \tan \alpha/\partial \theta = 0$, which leads to
\begin{equation}\label{rw1}
V_g^2 = V \frac{\partial V_g}{\partial \theta}\sin \theta + V V_g \cos \theta.
\end{equation}
The angle $\theta$ obtained from the above equation marks the point of intersection and, according to the Fig. \ref{fig:2}, describes the direction of radiation forming the cone. As it was shown in \cite{Carusotto} at this angle, the maximum of the radiation maximum intensity is expected. This is due to the fact that for all angles near $\theta$, the cone angle $\alpha$ is almost the same, so that constructive interference occurs.
In a dispersionless medium (e. g. $V_g=V_p$ and $\partial V_g/\partial \theta = 0$), the Eq. (\ref{rw1}) reduces to the classic relation for the Cherenkov radiation angle
\begin{equation}\label{rw2}
\cos \theta = \frac{V_p}{V}.
\end{equation}
In the case of a dispersive material, the Eq. (\ref{rw1}) can be solved separately for the two modes $V_{gi}(\theta)$, yielding characteristic angles $\theta_i$. These critical angles, as well as and the associated cones described by Eq. (\ref{rw0}) are shown on the Fig. \ref{fig:3a}. One can see that the first, wide cone is formed by the high frequency modes $\omega_1$. The radiation is emitted at an angle $\theta_1<\pi/2$ and the wavefronts are roughly parallel to the cone surface, in a manner similar to the Cherenkov radiation (Fig. \ref{fig:3b}). The low frequency modes $\omega_2$ form a much narrower cone. The emission angle $\theta_2 > \pi/2$ and the wavefronts are almost perpendicular to the cone surface which is a characteristic feature of the reversed Cherenkov effect \cite{Chen}. The first contour $V_{g1}(\theta)$ becomes almost circular for $\theta>\theta_1$. In this area, the radiated field is very similar to the case of the Doppler effect in dispersionless medium - a detector positioned at an angle $\theta$ will register only single wave mode, so that the wave frequency $\omega(\theta)$ can be readily measured.
On the other hand, as the angle approaches $\theta_0$, the frequency of the both modes changes quickly, so the field measured along some finite angle range around $\theta_0$ has a wide spectrum.
Further analysis of this case is presented on the Fig. \ref{fig:4}. The field snapshot from the FDTD simulation is presented in Fig. \ref{fig:4a}. Clearly, the obtained cones divide the radiated field into three distinct regions. The highest field amplitude is generated by the modes traveling along the direction $\theta_1$ and forming the outer cone. The inner cone is formed by the $\omega_2(\theta_2)$ modes. Another  significant direction is $\theta_0$, where the field is a superposition of many modes with a wide range of group velocities, so that it is spreading over time. These findings are confirmed by the calculated time-averaged value of the Poynting vector $\vec S = \vec E \times \vec H$ shown on the Fig. \ref{fig:4b}. One can see that the initial radiation pattern shows a wide peak at $\theta \approx \theta_1$ which is formed by the fast modes $\omega_1$. After the time needed for these modes to leave the simulation area, the intensity peaks formed by the slow modes $\omega_2$ can be observed. They are centered around $\theta_0$ and $\theta_2$. As it is shown on the Fig. \ref{fig:4c}, there is significant difference in the absolute value of the phase velocity between the fast modes $\omega_1$ and slow modes $\omega_2$. In the latter case, it is significantly smaller than $V$. Finally, the measured frequencies presented in Fig. \ref{fig:4d} are also in a good agreement with the predicted value. To illustrate the influence of the source velocity on the field structure, another simulation was performed for $\beta=0.1$. The results are shown on  Figs. \ref{fig:4e} and \ref{fig:4f}. In this case, the group velocity of the first mode $\omega_1$ is always significantly higher than $V$, so that no cone is formed and the radiation pattern differs only slightly from the field of radiating, stationary dipole. The second mode $\omega_2$ is characterized by a very low group velocity, forming a very narrow cone, barely visible in the field snapshot in Fig. \ref{fig:4e}.

\begin{figure}
\begin{minipage}[b]{.5\linewidth}
\centering
\includegraphics[scale=0.33]{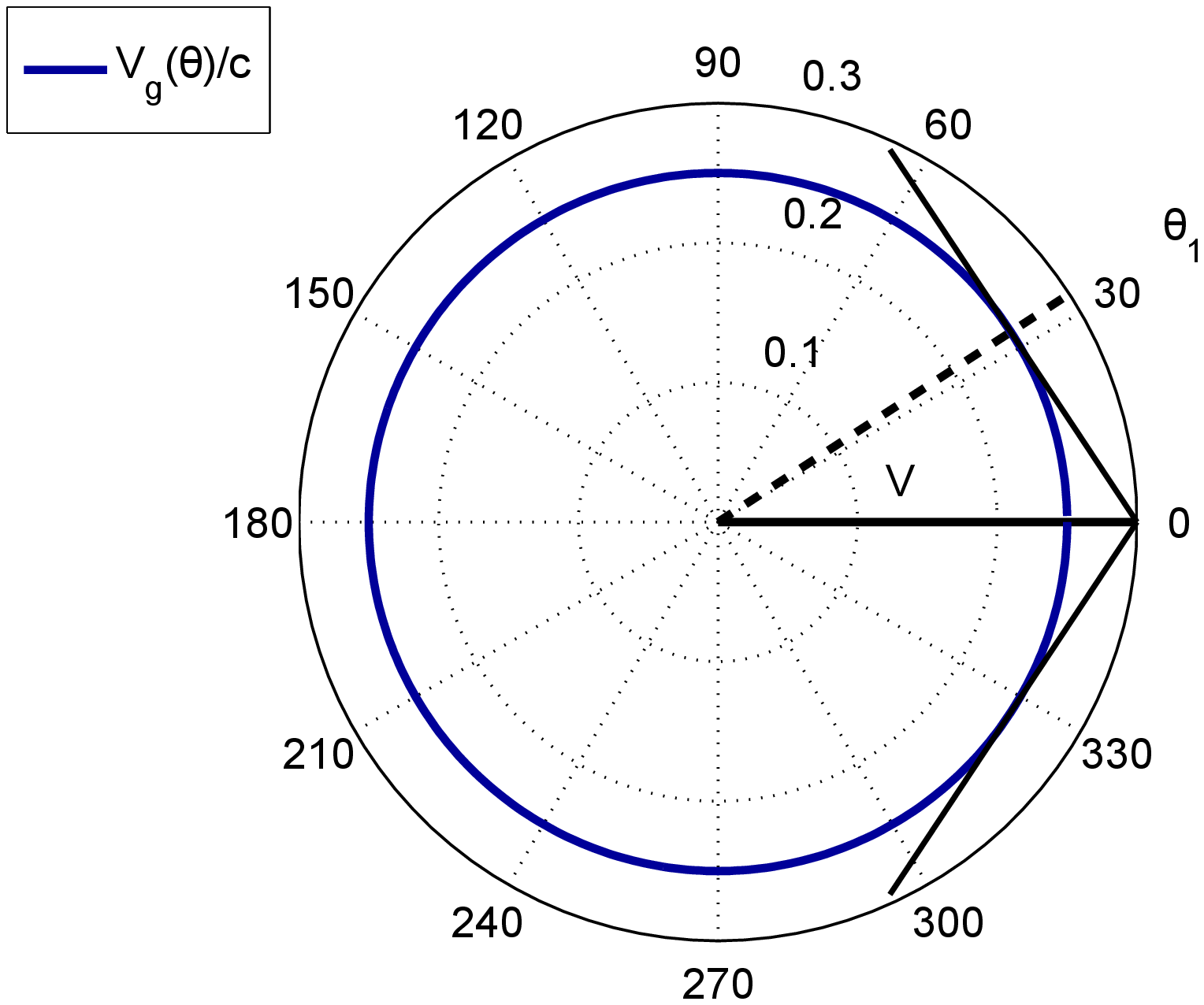}
\subcaption{}\label{fig:6a}
\end{minipage}%
\begin{minipage}[b]{.5\linewidth}
\centering
\includegraphics[scale=0.21]{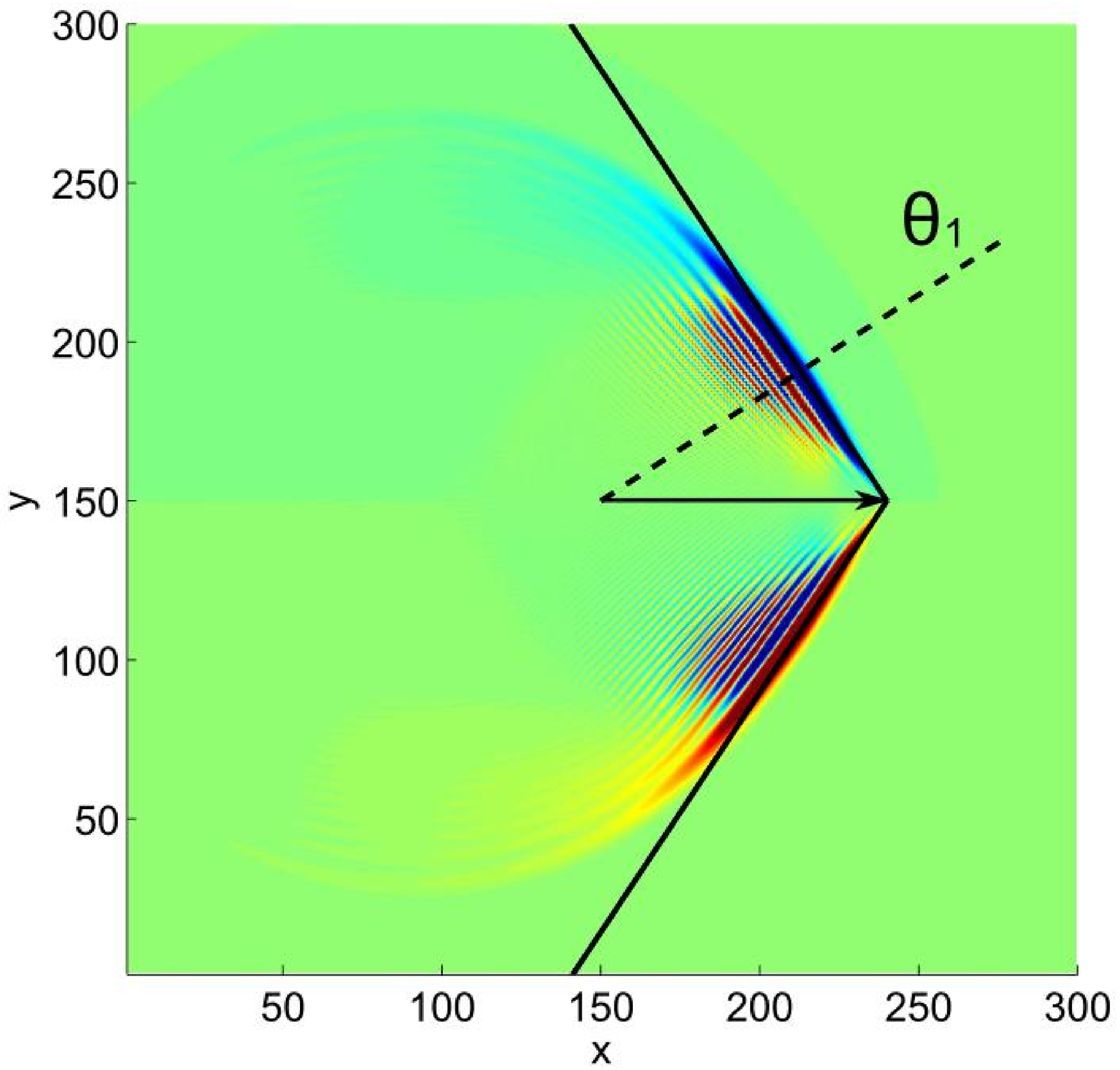}
\subcaption{}\label{fig:6b}
\end{minipage}

\begin{minipage}[b]{.5\linewidth}
\centering
\includegraphics[scale=0.33]{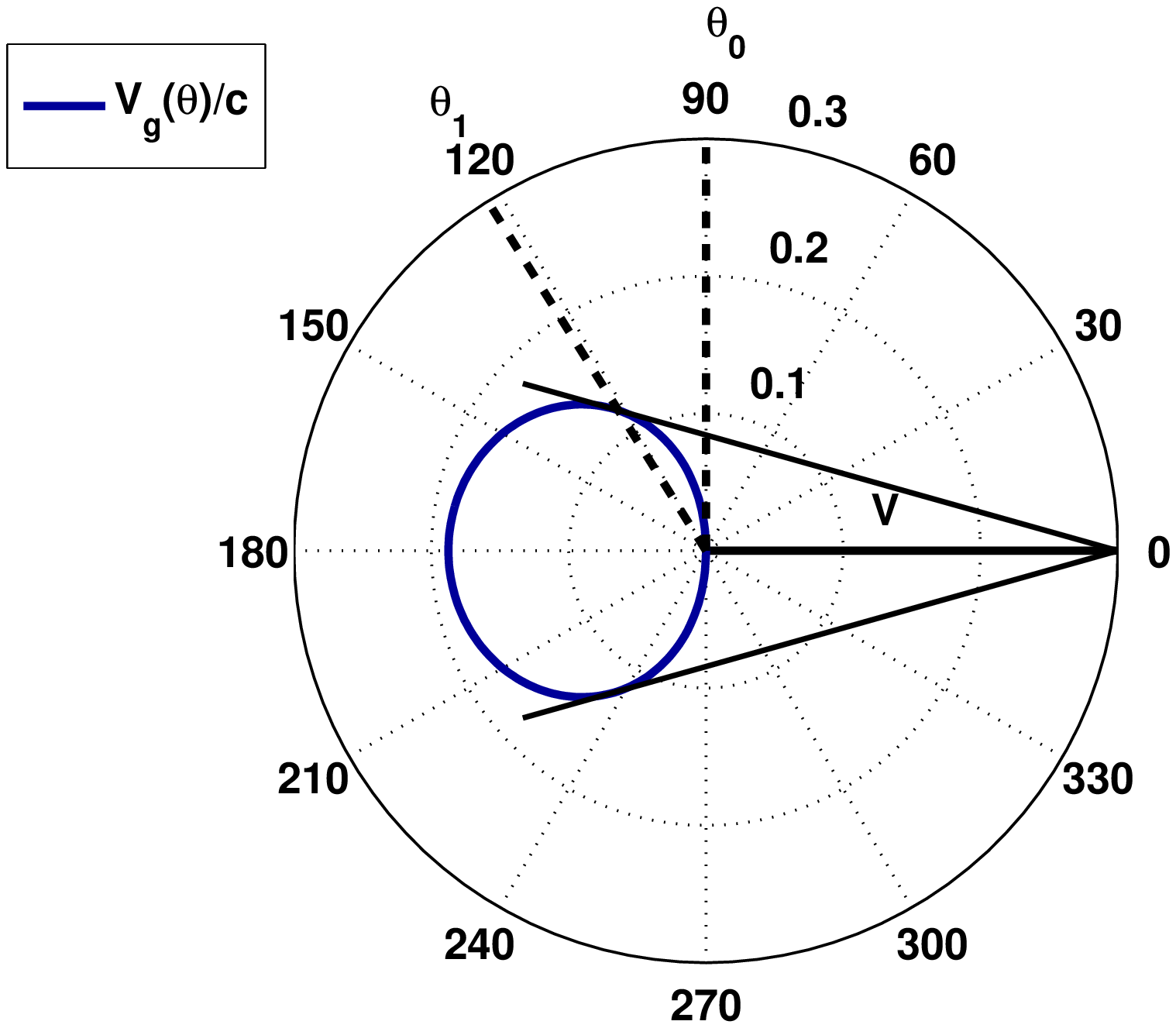}
\subcaption{}\label{fig:6c}
\end{minipage}%
\begin{minipage}[b]{.5\linewidth}
\centering
\includegraphics[scale=0.21]{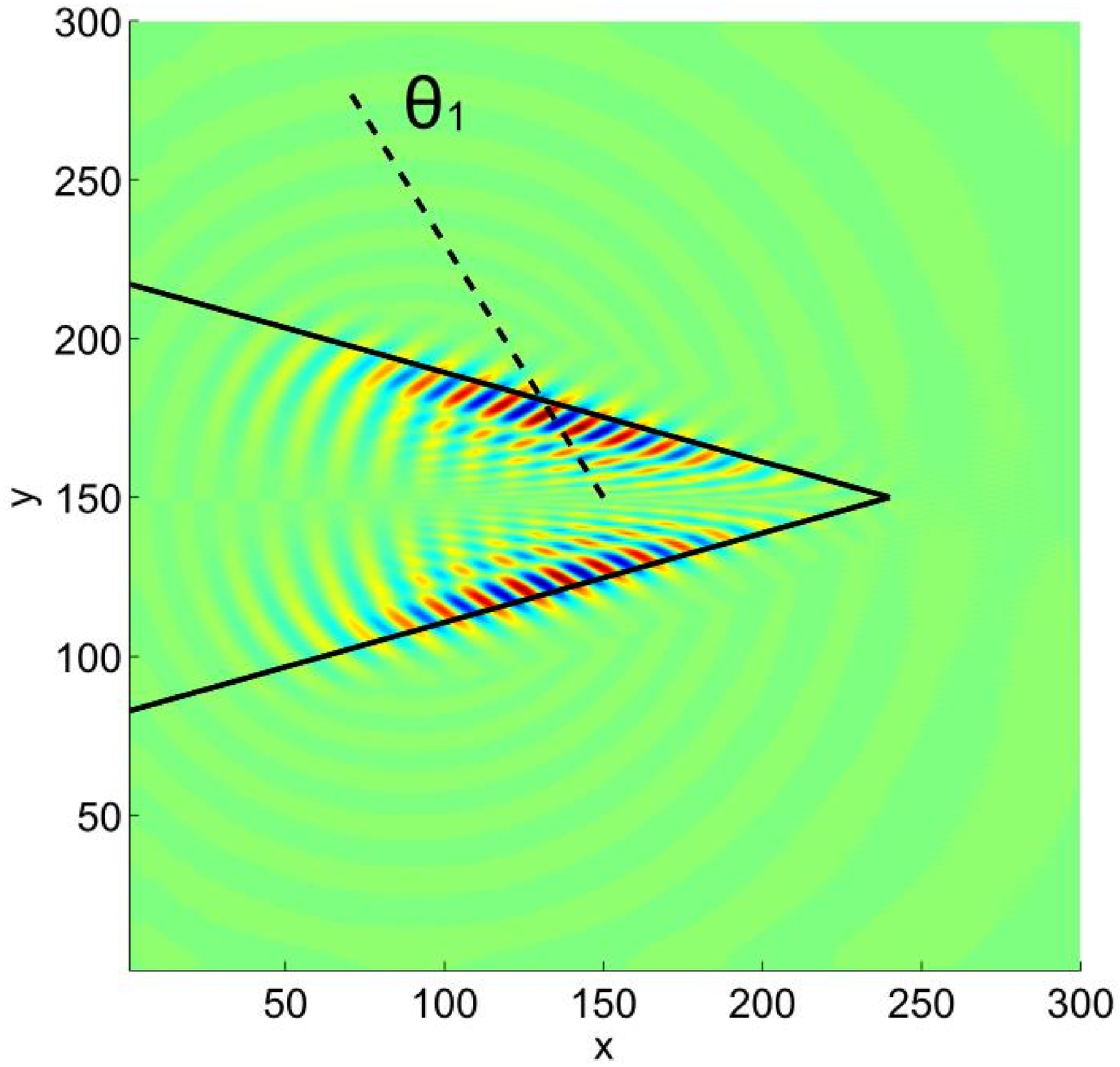}
\subcaption{}\label{fig:6d}
\end{minipage}

\begin{minipage}[b]{.5\linewidth}
\centering
\includegraphics[scale=0.33]{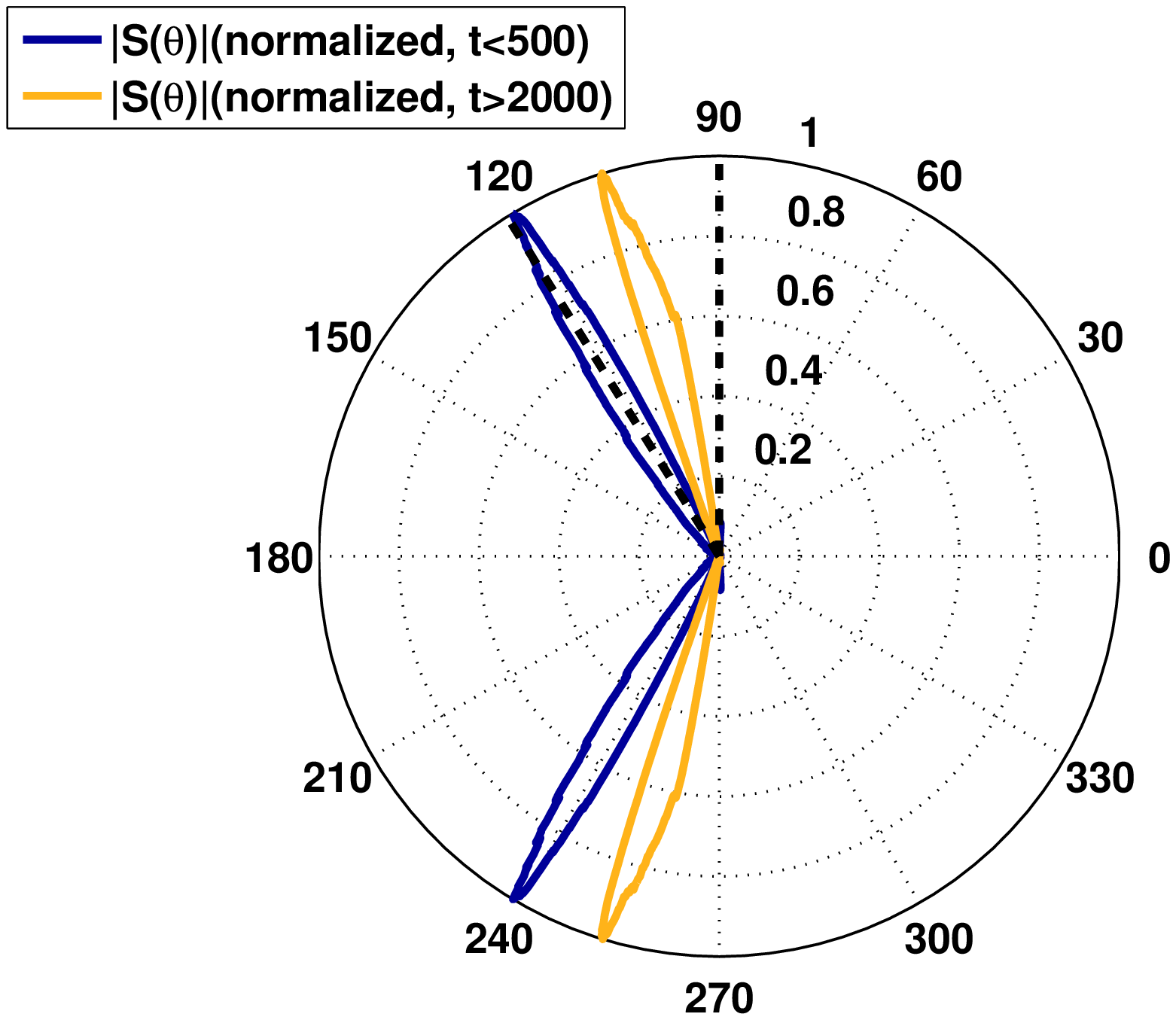}
\subcaption{}\label{fig:6e}
\end{minipage}%
\begin{minipage}[b]{.5\linewidth}
\centering
\includegraphics[scale=0.33]{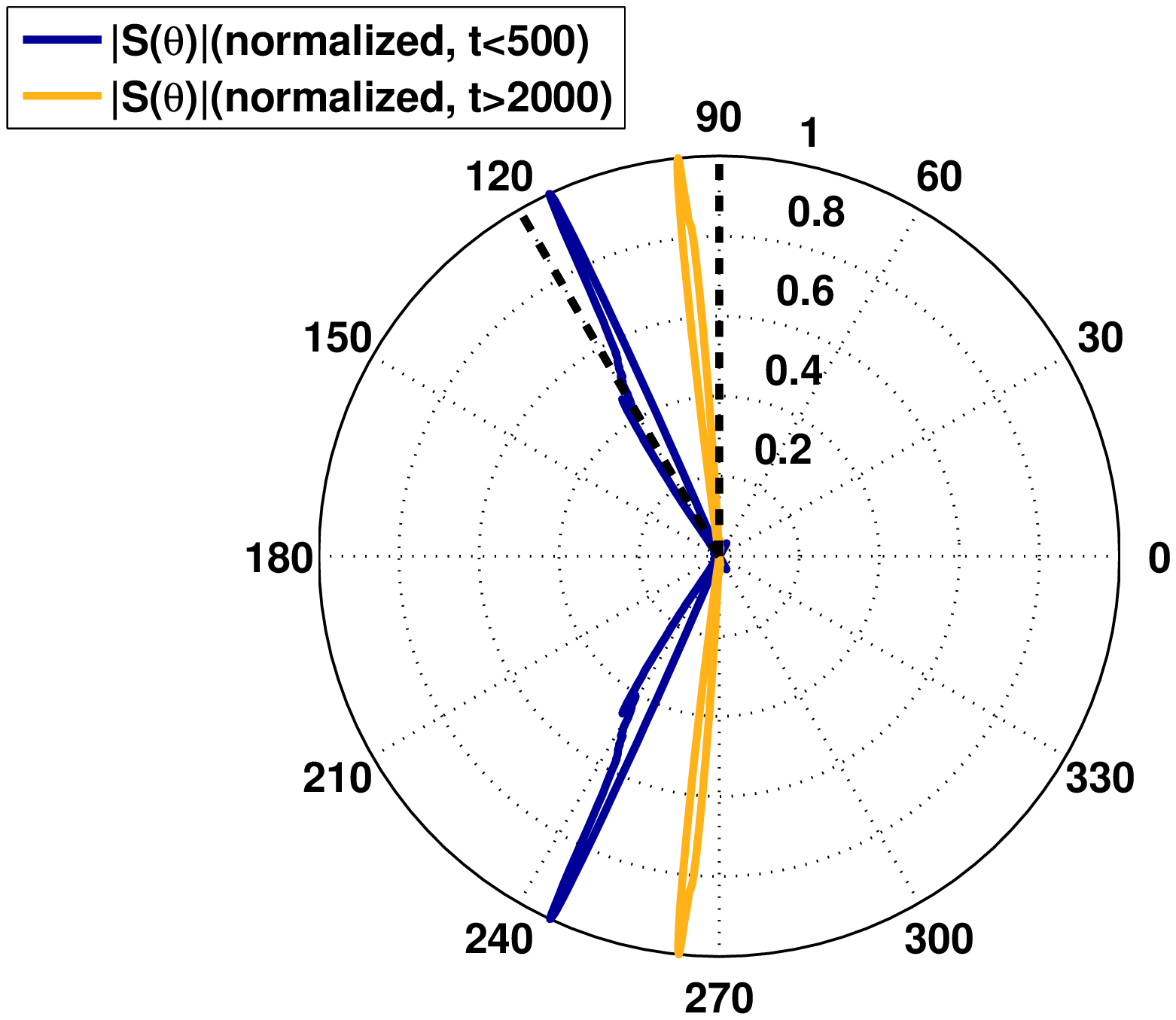}
\subcaption{}\label{fig:6f}
\end{minipage}

\caption{Cherenkov radiation for $\beta = 0.3$. (a) Group velocity contour in dispersionless dielectric with $n=4$. (b) Field snapshot at t=500, showing typical Cherenkov cone. (c) Group velocity contour in left-handed medium. (d) Field snapshot at t=500, showing reversed Cherenkov effect. (e) Radiation pattern for $\beta = 0.3$, exhibiting a narrow peak at $\theta_1$. (f) Radiation pattern for $\beta = 0.8$.}\label{fig:6}
\end{figure}

It is important to note the difference between the obtained radiation patterns and the Cherenkov effect. Here, the source of radiation is characterized by the nominal frequency $\omega_0$. Moreover, the absolute value of the phase velocity of the first mode is significantly larger than $V$ (Fig. \ref{fig:4c}), so that the condition in Eq. (\ref{rw2}) cannot be met. One can also see that the cone angle $\theta_1 < \pi/2$ which characterizes a non-reversed Cherenkov effect, despite the fact that the phase velocity is negative. Finally, for $\theta>\theta_1$, the cone transforms in a continuous manner into the typical Doppler shifted field pattern, generated by radiating point source. On the other hand, the second, narrower cone is formed by the modes characterized by $|V_{p2}| < V$ and $V_{g2} < V$; the radiation pattern is simpler, closely resembling the case of the reversed Cherenkov effect.

To compare the results to the generic Cherenkov radiation, our source model can be adapted to simulate a moving charge by assuming $\omega_0=0$. Under such condition, the  Eq. (\ref{DOPP1}) describes wave modes with frequency $\omega=\vec k \cdot \vec V = n \beta \cos \theta$, generating the spectrum of a point charge moving with velocity $V$. A particle of small but finite dimension moving along x axis can be modeled as a charge density \cite{Burlak}
\begin{equation}
\rho(x,y,t) = \rho_0 \exp\left[\frac{-(x-Vt)^2}{2\sigma^2}\right]\exp\left(\frac{-y^2}{2\sigma^2}\right),
\end{equation}
which corresponds to the source current \cite{Lu}
\begin{equation}
J = \rho_0V \exp\left[\frac{-(x-Vt)^2}{2\sigma^2}\right]\exp\left(\frac{-y^2}{2\sigma^2}\right).
\end{equation}
As a test case, a simulation was performed in a dispersionless dielectric characterized by $n=4$. The charge moving at a velocity $V=0.3~c$ is expected to create a cone at the angle $\theta_1 = \arccos (V_p/V) \approx 34^o$. The geometrical construction based on the group velocity contour is presented on the Fig. \ref{fig:6a}. The field snapshot is shown on the Fig. \ref{fig:6b}. The Cherenkov cone is clearly visible and its angle matches the theoretical prediction.
In the case of a dispersive medium described by Eq. (\ref{modDRU}), the Doppler shift Eqs. (\ref{DOPP2}) remain valid for $\omega_0=0$, reducing to a single solution
\begin{equation}\label{DOPP5}
\omega_1(\beta,\theta) = \omega_2(\beta,\theta)= \omega_p\sqrt{\frac{-\beta\cos\theta}{1-\beta\cos\theta}}.
\end{equation}
One can check that for any angle $\theta$, the phase velocities of the wave modes described by above relation fulfill the Eq. (\ref{rw2}). This important property should be stressed; in the classical Cherenkov radiation, the single radiation angle is a function of the two constants - the phase velocity and the source velocity. Here, the phase velocity of the emitted mode varying, being a function of the source speed and angle $\theta$. In other words, the radiation pattern is a superposition of cones corresponding to different group velocities \cite{Luo}. The real value of the frequency associated with propagating waves is obtained from Eq. (\ref{DOPP5}) only for $\theta>\theta_0=\pi/2$. This means that the Cherenkov effect is reversed which is consistent with the negative value of the refraction index \cite{Veselago}. The group velocity contour for $V=0.3c$ is shown on the Fig. \ref{fig:6c}. In contrast to the Cherenkov effect in dispersionless medium, the frequency range of the emitted modes is limited, and their group velocities vary from $V_g=0$ at $\theta=90^0$ to $V_g \approx 0.19~c$ at $\theta=180^0$. Due to the low values of the group velocity, the outer cone is relatively narrow, and the radiation angle $\theta_1 \approx 120^o$. Again, the structure of the field in the snapshot presented in Fig. \ref{fig:6d} is in an excellent agreement with the analytical prediction. As expected, the radiation pattern presented in Fig. \ref{fig:6e} shows a narrow initial peak at $\theta=\theta_1$, followed by the slow modes at $\theta_0>\theta>\theta_1$.
When a single frequency is measured, the radiation angle agrees with the generic Cherenkov condition given by Eq. (\ref{rw2}) for the appropriate value of $n(\omega)$. Therefore, the performed simulations show a good agreement with prior simulations where a window function has been used to reduce frequency range \cite{Chen} and with experimental data obtained by Xi \emph{et al.}\cite{Xi}.

\begin{figure}
\begin{minipage}[b]{.5\linewidth}
\centering
\includegraphics[scale=0.25]{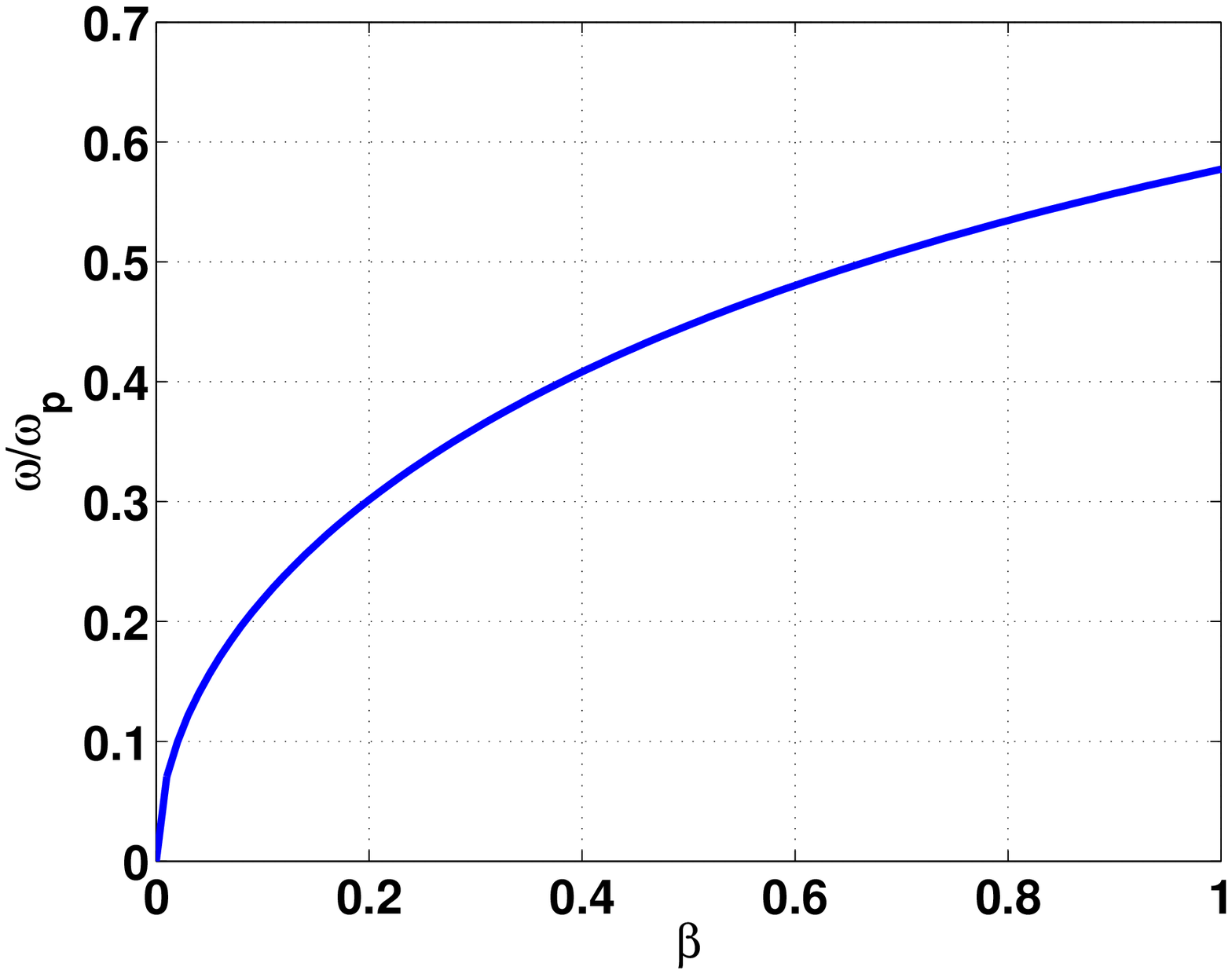}
\subcaption{}\label{fig:7a}
\end{minipage}%
\begin{minipage}[b]{.5\linewidth}
\centering
\includegraphics[scale=0.25]{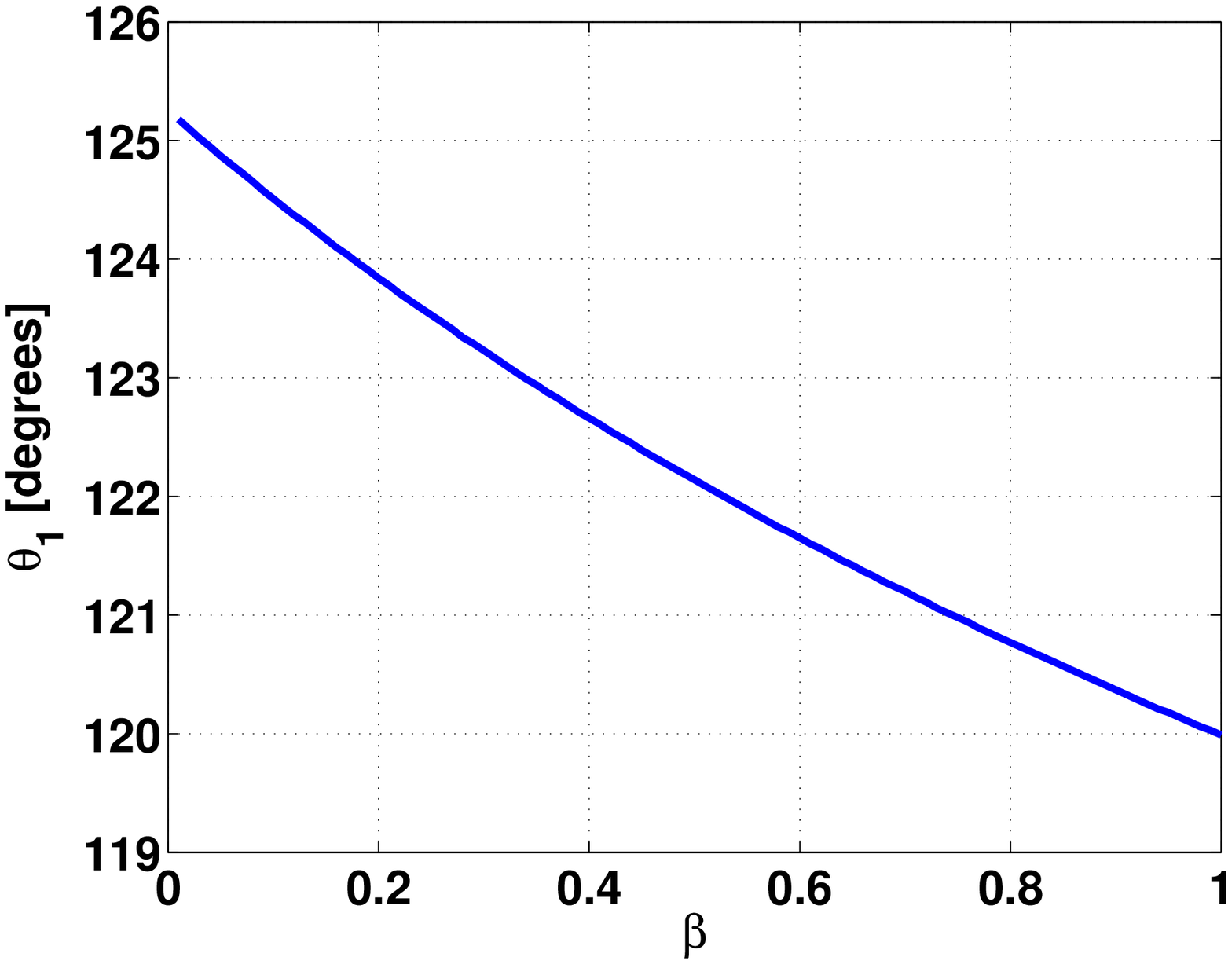}
\subcaption{}\label{fig:7b}
\end{minipage}

\caption{(a) The dependence of the cone angle on the source velocity. (b) The dependence of the wave frequency on the source velocity for $\theta=120^0$.}\label{fig:7}
\end{figure}

Surprisingly, the dependence of the cone angle on the source velocity is insignificant; the result for $V=0.8 c$ shown on the Fig. \ref{fig:6f} is almost indistinguishable from the case of $V=0.3 c$. This is confirmed on the Fig. \ref{fig:7a}, where the numerically computed relation $\theta_1(\beta)$ based on Eq. (\ref{rw1}) is shown. This interesting property depends on the dispersion relation of the medium and has been observed  in various physical systems;  for example, in the so-called Kelvin wake in water waves \cite{Georgi}, the group velocity is directly proportional to the phase velocity, which is given by Eq. (\ref{rw2}). This causes the Eq. (\ref{rw1}) to independent of the source speed; the details are presented in Appendix A.

One can see another significant departure from the typical Cherenkov effect in a dispersionless medium; the Eq. (\ref{DOPP5}) has a solution for any $1 \geq \beta \geq 0$. Therefore, there is no source velocity threshold above which the phenomenon occurs. Unless additional constraints are put on the dispersion model, the frequency and phase velocity of the wave modes can become arbitrarily small, matching the source velocity in accordance with the Eq. (\ref{rw2}) and producing radiation at any $\beta$. Such radiation without a velocity threshold has been observed with metallic gratings \cite{Purcell} and predicted to occur in periodic structures \cite{Luo} such as nanowire metamaterial \cite{Fernandes}. The Cherenkov radiation frequency varies considerably with  source velocity (Fig. \ref{fig:7b}) but, in general, it is comparable to the resonant frequency of the medium $\omega_p$, so that it is defined by the material properties \cite{Burlak}.

\section{Conclusion}

We presented an unified analytical description of the two-dimensonal Doppler  and Cherenkov effects in an idealized, dispersive metamaterial.  It was shown that a moving, monochromatic source generates two distinct frequency modes and the generated field exhibits features of both, the Doppler and the Cherenkov effects. We prove,
starting from the first principles, that the reversed Cherenkov radiation emerges as a particular case of the presented theory.  The characteristic peculiarities
of the reversed Cherenkov effect in LHM, namely the backward direction of emission, the constant angle of maximum radiation intensity and the lack of a source velocity threshold above which this phenomenon occurs are explained on the basis of  presented theory and confirmed by numerical simulations. Our results are in agreement with measurements obtained in the experiment performed by Xi \emph{et al.}\cite{Xi}. All these interesting features of the Cherenkov radiation in metamaterials characterized by Drude-like dispersion model might possibly provide a new way of frequency-based charged particle speed measurement, with the mechanical simplicity enabled by the constant radiation angle.

\newpage
\appendix
\section{The angle of Cherenkov radiation} \label{App:AppendixB}
\begin{figure}[ht!]
\begin{center}
    \includegraphics[scale=0.35]{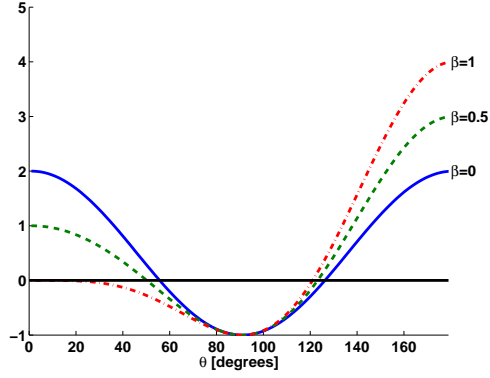}
  \caption{Plot of Eq. \ref{rwA1} for selected values of $\beta$.}\label{fig:A1}
\end{center}
\end{figure}

In the Drude model medium described by Eq. (\ref{modDRU}), the refraction indices for the phase and group
velocities are given by
\begin{equation}
n = 1-\frac{\omega_p^2}{\omega^2}\quad n_g = 1+\frac{\omega_p^2}{\omega^2},
\end{equation}
therefore, these velocities are linked by the relation
\begin{equation}\label{R1}
V_g = \frac{V_p}{2 \frac{V_p}{c} - 1} = \frac{V \cos \theta}{2\beta\cos \theta - 1},
\end{equation}
where the phase velocity is defined by the generic Cherenkov radiation condition in Eq.~(\ref{rw2}). The Eq.~(\ref{rw1}) defining the angle of maximum radiation intensity can be simplified to the following form
\begin{equation} \label{rwA1}
\cos^2\theta(3 - 2\beta \cos \theta) -1 = 0.
\end{equation}
Considering the fact that the real frequency modes are given by Eq. (\ref{DOPP5}) only at angles bigger than $90^o$, for any given source speed $\beta$, the above relation has one proper solution at $\theta \approx 120^0$ (Fig. \ref{fig:A1}). One can check that in the limit of $\beta \rightarrow 0$, $\theta \approx 125^o$ and for $\beta = 1$, $\theta = 120^o$.

It is interesting to mention that in the case of the water waves, one has \cite{Georgi}
\begin{equation}
V_g = \frac{1}{2}V_p = \frac{V \cos \theta}{2}.
\end{equation}
With such defined group velocity, the Eq. (\ref{rw1}) takes a simple form
\begin{equation}
\sin^2 \theta = \frac{1}{3},
\end{equation}
with a solution describing the well-known Kelvin wake angle $\theta \approx 35.26^o$.

\end{document}